# Multiplexed back focal plane imaging with on-chip integrated microlens array

M. Furman[1], M. Muszyński[1], P. Oliwa[1], Ł. Zinkiewicz[1], A. Bogucki[1,2], J. Szczytko[1], P. Wasylczyk[1], W. Pacuski[1], M. Król[1], B. Piętka[1]

[1] Institute of Experimental Physics, Faculty of Physics, University of Warsaw, ul. Pasteura 5, PL-02-093 Warsaw, Poland
[2] Center for Quantum Nanoscience, Institute for Basic Science, Seoul 03760, Republic of Korea
barbara.pietka@fuw.edu.pl

On-chip optical architectures that enable angle-resolved spectroscopy are essential for advancing photonic platforms towards low-volume, scalable, and cryo-compatible devices. Here, we introduce spatially resolved momentum-space imaging using arrays of 3D-printed microlenses directly integrated onto semiconductor optical microcavities. Each microlens functions as an independent optical element with a high numerical aperture (approx. 0.95), enabling parallel back focal plane imaging across, in our implementation, 64 distinct locations. This approach eliminates the need for bulky microscope objectives while maintaining broad wavevector access, even under cryogenic conditions. We demonstrate its versatility across various cavity systems, including dielectric planar resonators, GaAs-based polariton microcavities, and CdTe platforms supporting nonequilibrium Bose-Einstein condensation of exciton-polaritons. The microlenses not only enhance collection efficiency but also enable tightly focused excitation, yielding an order of magnitude reduction in condensation thresholds compared with conventional setups. Our results establish 3D-printed microlens arrays as a compact, versatile, integrated platform for next-generation angle-resolved spectroscopy in nanophotonics and quantum materials.

## Introduction

Back focal plane (BFP) imaging, also referred to as reciprocal space, $k$-space, or Fourier plane imaging, is a widely used experimental technique that maps the angular distribution of light emitted or transmitted by a sample. In many optical and photonic systems, this angle is directly linked to the in-plane wavevector (momentum) of light or quasiparticle excitations.

As such, momentum-resolved spectroscopy offers direct insights in all subfields of photonics[1,2]. Yet, accessing momentum-space typically relies on bulky microscope objectives that limit experimental flexibility, especially under cryogenic conditions or in scalable device geometries. On-chip optical elements offer a pathway toward compact and integrable solutions, where light can be collected, focused, and manipulated directly at the device level. In particular, three-dimensional printing of microoptical components provides remarkable freedom to design monolithic architectures with site-specific functionality.

In nanophotonics and plasmonics, the directionality of light propagation plays an essential role. There, BFP imaging has been adapted to investigate various devices such as plasmonic nanowires[3,4] or nanoantennas[5–7], as well as diverse optical phenomena such as surface-enhanced Raman scattering[8], second and third harmonic generation[9,10], cathodoluminescence[11,12], and up-conversion[13]. The technique has also proven valuable in the quantum regime[14–16], where it provides access to the angular emission characteristics of quantum emitters.

On the fundamental level, BFP imaging allows to determine the direction of a dipole transition, such as identifying the orientation of single molecules[17,18] or carbon nanotubes[19]. The same method has been used to distinguish the in-plane or out-of-plane character of excitons in layered materials[20,21], their heterostructures[22] and thin organic films in OLEDs[23]. It also allows to differentiate from the most common electric-dipole transitions scarce magnetic-dipole ones, occurring in rare earths complexes[24] or in 2D perovskites[25].

In systems such as photonic crystals, nanostructures, metasurfaces, or optical cavities, the wavelength and wavevector of photons are connected, forming intricate energy-momentum dispersion relations. There, momentum-resolved spectroscopy enables the experimental study of a wide range of physical phenomena originating from the specific structure in momentum-space. This includes topological states, such as bound states in the continuum[26,27], topologically protected edge modes[28–30], and photonic analogues of electronic topological insulators[31,32]. It also allows exploration of solid-state analogies, including photonic lattices[33], Dirac cones[34], and Berry curvature-related phenomena[35–37]. Furthermore, this approach provides access to non-Hermitian physics, such as manipulation of exceptional points[38,39], or optical simulations of exotic physics far from classical optics, such as analogues of black hole horizons[40,41] and new phases of matter[42].

Conventional approaches to reciprocal space imaging rely on high-magnification, high-numerical-aperture (NA) microscope objectives to access a broad angle range, often spanning several tens of degrees. These same objectives are typically used for excitation, as their large NA enables the tight focusing required to achieve the high power densities necessary for reaching nonlinear phenomena. However, these optical elements come with significant limitations. Their short working distances make integration with cryogenic systems particularly challenging. Moreover, they inherently restrict measurements to a single spatial location at a time, since they integrate angular information over the entire collection area. As a result, conventional excitation and detection schemes lack the capability for spatially multiplexed momentum-space imaging, limiting experimental access to non-local phenomena. This precludes studies of spatially distributed momentum-space dynamics and related phenomena such as Josephson oscillations, site-to-site couplings, or momentum-dependent particle population, flow, and thermalization.

Here, we present a novel photonic approach based on ellipsoidal microlenses directly printed on top of optical microcavities, replacing the conventional microscope objective in both the excitation and detection paths. This allows for simultaneous momentum-space imaging of multiple (up to 64) spatial positions, each with access to large emission angles (up to 72 degrees in air, NA = 0.95), all while eliminating the need for bulky microscope objectives. The high NA of microlenses also provides tight focusing of the excitation beam, which significantly reduces the threshold of polariton Bose-Einstein condensation by nearly an order of magnitude compared to conventional microscope objective excitation. The versatility and robustness of printed microlenses is demonstrated in both purely photonic and exciton-polariton optical microcavities.

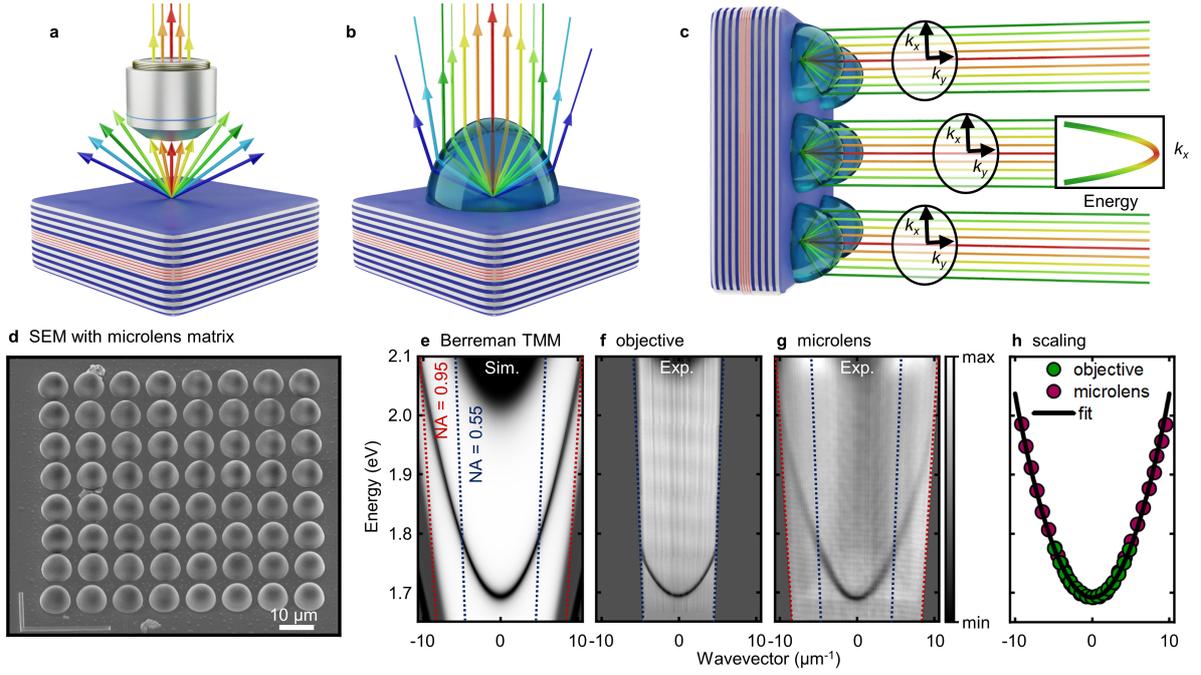

**Fig. 1 Ellipsoidal microlenses printed on the surface of an optical microcavity, serving as miniaturized detection optics to observe the dispersion relation over a broad range of in-plane wavevectors (angles).** Schematic illustration of rays of light emitted from a structure at different angles (marked with colors) collected using **a**, a microscope objective with high numerical aperture, and **b**, a microlens printed directly on top of the sample. **c**, Schematic of the experimental configuration where an array of printed ellipsoidal microlenses acts as a set of independent, miniaturized optical systems with an exceptionally wide collection angle. This setup enables simultaneous back focal plane imaging of multiple spatial positions. **d**, SEM image of an 8×8 square array of microlenses printed on the surface of the microcavity. **e**, Simulated momentum-resolved reflectivity spectrum of a $SiO_2/TiO_2$ dielectric photonic microcavity, calculated using the Berreman method. Dashed lines indicate the wavevector ranges corresponding to the numerical apertures of a standard objective (navy blue, NA = 0.55) and a microlens (dark red, NA = 0.95). **f,g**, Experimental momentum-resolved white-light reflectivity spectra of the same microcavity, collected using (**f**) a conventional objective (NA = 0.55), and (**g**) a printed microlens (NA = 0.95). **h**, Extracted resonant energies from spectra in panels (**f,g**). Dark green points correspond to data from panel (**f**), the black curve shows a fit to the dark green points, and dark pink points correspond to data from panel (**g**).

## Experimental results

The light ray paths in the traditional configuration with an optical microscope and presented here microlenses are shown in Figs. 1**a**,**b**, respectively. In the microscopic configuration (Fig. 1**a**), the collection of a high emission angles is limited by the geometry of the objective, its NA and working distance. On the other hand, a microlens (Fig. 1**b**) located directly on the surface of the sample, thanks to its solid immersion properties, allows for more effective light harvesting, especially at high emission angles, which correspond to a higher NA.

Polymer, micrometer-sized lenses have already been used for efficient objective-free light collection[43]. However, that design optimized for emitted beam collimation results in a highly distorted Fourier plane, hin-dering momentum-resolved imaging. Here, we developed ray-tracing simulations [see Supplementary Information (SI), section Ray Tracing] to design a microlens with optimized shapes of the back focal plane. The goal was to maximize the angle of light collection (targeting a numerical aperture of up to NA ≈ 0.95) while maintaining high fidelity of the Fourier plane by minimizing optical aberrations and distortions. In addition, the geometry of the microlenses was optimized to precisely match the structure of each sample.

The ellipsoidal microlenses were fabricated by two-photon polymerization lithography (for more details, see Methods, section Two-photon Lithography). Their final dimensions, namely the vertical half-axis $a$, the horizontal half-axis $b$, and the cutting position $d$ along the $z$-axis, are provided for each structure in Table I of the SI. The Fourier plane corresponding to each microlens is located approximately 10–15 μm above the top of the microlens. An optical setup allowing imaging of this Fourier plane projected by the microlens and performing momentum-resolved measurements is presented in Fig. S5**b** in SI. This configuration is analogous to a conventional microscope, where a microscope objective and an imaging (tube) lens project a magnified image of the Fourier plane of the microlens onto a camera or the entrance slit of a spectrometer.

However, the use of microlenses provides an additional advantage. In a conventional $k$-space imaging setup, access to the Fourier plane is limited to a single spatial position at a time and to light rays that have not diverged significantly before entering the microscope objective. In contrast, our approach replaces the objective with microlenses directly fabricated on the sample surface, which locally collimate the emitted light. Thus, each microlens acts as an independent miniature optical system, enabling parallel momentum-space mapping across a spatially multiplexed array of emission sites, as shown in Fig. 1c.

To demonstrate this use case, we fabricated a lattice of ellipsoidal microlenses arranged in a regular 8×8 square array with a center-to-center spacing of 20 μm, providing sufficient spatial separation to prevent optical cross-talk between neighboring elements, while allowing simultaneous acquisition of $k$-space information from all microlens positions. A representative scanning electron microscope image of the fabricated array is shown in Fig. 1d.

As an illustration of the versatility of our approach, we applied those structures to three distinct types of optical microcavity systems: (Sample A) a purely photonic $SiO_2/TiO_2$ microcavity operating at room temperature; (Sample B) a GaAs-based microcavity, operating at cryogenic temperatures in the strong light-matter coupling regime and characterized with homogeneous photonic potential; and (Sample C) a CdTe-based microcavity exhibiting nonequilibrium Bose-Einstein condensation within intrinsic photonic potential traps. As a consequence of strong coupling, exciton polaritons, mixed light–matter quasiparticles, are formed. They are characterized by two distinct dispersion branches (upper and lower polaritons), whose curvature depends on the detuning between the photonic mode and the exciton in the quantum well. Under strong and tightly focused excitation, these quasiparticles enable the study of nonlinear interactions above the threshold for a phase transition to a Bose-Einstein condensate.

We first characterized the optical signal collected through a single microlens at room temperature using Sample A, which consists of a simple and well-defined dielectric stack supporting a single photonic cavity mode (see Methods, section Samples, for structural details). To demonstrate the broad angle (wavevector) collection capability of our approach, Figs. 1e–g present a comparison between momentum-resolved reflectivity spectra obtained by simulation and experiment. Specifically, Fig. 1e shows the theoretically calculated dispersion using the Berreman method[44,45], while Figs. 1f,g show experimental spectra measured using a planar cavity in setup described in Fig. S5a in SI and with a single microlens and microscope objective 50× as a Bertrand lens in setup Fig. S5b in SI, respectively.

In all three cases, the parabolic dispersion of the cavity mode is clearly observed. In Fig. 1e, dashed lines mark the wavevector ranges corresponding to the numerical apertures of NA = 0.55 and NA = 0.95, which correspond well to the wavevector ranges observed experimentally in Figs. 1f,g, respectively.

To validate the wavevector range in the experimental data, the cavity mode dispersion was extracted by Lorentzian fitting to the resonance energies and compared to the simulated dispersion (see Fig. 1h). These results confirm that the microlens not only serves as a light collection element but also provides access to a significantly broader wavevector range, reaching NA ≈ 0.95.

Although under ambient conditions, high-NA microscope objectives offer similar performance to microlens, they become impractical for use in cryogenic experiments. This limitation is particularly relevant for studies of semiconductor-based polaritonic microcavities, where experiments are typically performed at low temperatures using closed-cycle or liquid helium cryostats. In contrast, microlenses can be directly integrated into the cryogenic environment and cooled along with the sample. This approach enables broad wavevector-range momentum-space imaging at cryogenic temperatures, without relying on bulky, high-NA microscope objectives in the detection path.

Figure 2 presents a comparison of momentum-resolved photoluminescence spectra acquired from strongly-coupled semiconductor cavity, Sample B, using setup from Fig. S5b in SI (i.e., with a single microlens) in combination with various Bertrand lenses. Microcavity exhibits clearly resolved upper and lower polariton branches. The observed polariton dispersion was fitted using a coupled oscillator model and superimposed on the experimental spectra for reference (for details, see SI, section Additional Experimental Results).

The measurements shown in Figs. 2a–e were obtained at 4.5 K using Bertrand lenses of varying focal lengths: microscope objectives with $f = 4$ mm (50×), $f = 10$ mm (20×), $f = 20$ mm (10×), and $f = 40$ mm (5×), as well as a $f = 25.4$ mm biconvex lens, respectively, for both excitation and collection of light. All spectra exhibit a qualitatively consistent cavity polariton dispersion, with typical anticrossing behavior of the upper and lower polariton branches. The variation of Bertrand lenses affects the beam spot size on the microlens under fixed excitation power, thereby influencing the local excitation density, which leads to a modified population distribution within the lower polariton branch.

Figure S10 in the SI section Additional Experimental Results presents a comparison of measurements on Sample B taken without and with a single microlens. This reveals an increased observable wavevector range up to 7.13 μm$^{-1}$, in excellent agreement with the target NA = 0.95.

Local strain induced by the microlens printing process results in a blueshift of both the exciton and cavity mode energies. This effect can be seen, for example, in Fig. 2d, where the signal at 1.457 eV corresponds to the minimum of the lower polariton branch in the unstrained region outside the area covered by the microlens and was taken into account in the modeling (Additional Experimental Results section in SI).

As the magnification of the optics in the detection decreases, a broadening of the polariton branches is observed. This is attributed to the reduced wavevector resolution per pixel, caused by the smaller size of the Fourier plane image projected onto the detector. Although this loss of resolution can be compensated by introducing an additional telescopic magnification element in the detection path, this was omitted to maintain consistent experimental conditions. In the case of the simple biconvex lens, the additional spectral broadening is due to the lack of spherical aberration correction, which is otherwise provided by high-resolution microscope objectives.

Nevertheless, these results clearly demonstrate that momentum-resolved polariton spectroscopy can be effectively performed using only the microlens in combination with low-NA optics, thereby eliminating the need for conventional microscope objectives in the detection path, even under cryogenic conditions.

The additional advantage of a microlens approach is ability to multiplex momentum-space measurements from a microlens array. Fig. 3 demonstrates the simultaneous momentum-space imaging of exciton-polariton photoluminescence from multiple spatially separated positions on Sample B. This is made possible by the use of multiple independent microlenses, each acting as an independent optical system. The entire 8×8 microlens array was illuminated with an expanded laser beam, using a 10× microscope objective ($f = 20$ mm) as the Bertrand lens. Equivalent measurements using a $f = 25.4$ mm focal length lens are provided in the SI (see section Additional Experimental Results, Fig. S11).

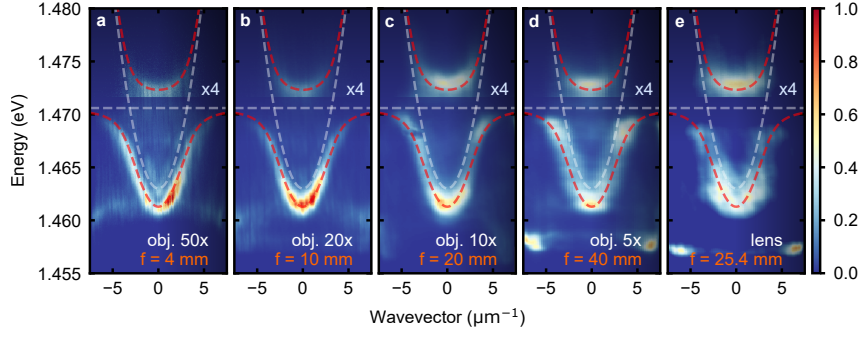

**Fig. 2 Exciton-polariton dispersion relation from a GaAs semiconductor optical microcavity collected with an ellipsoidal microlens and different Bertrand lenses. a–e,** Momentum-resolved photoluminescence spectra collected using an ellipsoidal microlens printed directly on the sample surface. After passing through the microlens, the signal was magnified and directed into the spectrometer slit using either short focal length lens or microscope objectives. The specific objective or lens used in each measurement is indicated in the lower right corner of each panel. The polariton dispersion was modeled using the coupled oscillator model (red dashed lines), while the uncoupled photon and exciton modes are shown as white dashed lines. The signal intensity from the upper polariton branch was multiplied to enhance visibility. All data were acquired at 4.5 K in a cryostat.

Figure 3a shows the multiple momentum-resolved photoluminescence spectra acquired in a single measurement by projecting Fourier planes from a row of eight microlenses onto the entrance slit of the spectrometer. Each microlens provides a well-resolved polariton dispersion, with clearly separated upper and lower branches. To achieve spatially varying population distributions within the lower polariton branch, we employed an excitation beam with a Gaussian intensity profile rather than a flat-top distribution. This resulted in a non-uniform excitation power density across the illuminated microlens array. Consequently, microlenses located near the center of the excitation spot exhibit enhanced emission at the minimum of the lower polariton branch, whereas those positioned toward the edges show a more uniformly populated dispersion.

Furthermore, momentum-space maps spectrally filtered at selected photoluminescence energies and collected on the CCD camera, as shown in Figs. 3b–d, provide isoenergetic momentum-space cross-sections simultaneously captured from all 64 microlens positions. These energy slices correspond to different parts of the lower polariton branch, namely, the high in-plane momentum region, the mid-dispersion region, and the branch minimum. The specific energies at which the slices were collected are indicated with colored dashed lines in Fig. 3a for reference.

These results demonstrate that even with a relatively simple optical experiment, the printed microlens array enables parallel, spatially resolved momentum-space spectroscopy, allowing simultaneous investigation of polariton dispersion and population dynamics at multiple distinct positions across the sample.

The next step of our study aimed to demonstrate the influence of microlens-assisted excitation on nonequi-librium Bose-Einstein condensation and associated threshold pulse energies. To achieve this, arrays of microlenses were printed on Sample C, which consisted of a CdTe/CdMgTe microcavity embedding 6 quantum wells. This structure features intrinsic photonic disorder that forms natural photonic traps where spatially localized polariton condensation typically occurs[46]. In this case, the microlens shape was optimized to achieve an even higher NA = 1, at the cost of a slightly distorted dispersion image (see Fig. S2a).

Figures 4a,b show momentum-resolved photoluminescence spectra acquired in the setup presented in Fig. S5b, collected below and above the condensation threshold, respectively. Below the threshold, the spontaneous emission is uniformly distributed along the lower polariton branch. Above the threshold, a sharp and intense signal emerges, indicating that the condensate predominantly occupies localized trap states formed by intrinsic photonic disorder, behavior that is typical for CdTe-based microcavities[47].

Figures 4c,d present the emission intensity, spectral linewidth, and energy blueshift as a function of excitation power, measured under the microlens using setup from Fig. S5b, and in a nearby unpatterned region using typical imaging setup from Fig. S5a in SI with a 50× objective (NA = 0.55), respectively. In both cases, the data exhibit signatures of polariton condensation in semiconductor microcavities: a nonlinear increase in emission intensity, a pronounced narrowing of the spectral linewidth, and a blueshift of the emission energy at the condensation threshold.

However, a notable difference is observed in the excitation pulse energy required to reach the condensation threshold. With the microlens excitation, the threshold is nearly an order of magnitude lower compared to that in neighboring unpatterned regions. To systematically confirm this, two measurement series were performed for two different exciton-photon detuning (see SI, section Additional Experimental Results, Fig. S12 for data corresponding to the second detuning). Each series included measurements from 13 identical microlenses and 13 neighboring unpatterned areas. The resulting condensation threshold values are summarized in the histograms shown in Fig. 4e. Here, orange bars represent thresholds obtained under microlens excitation, while royal blue bars correspond to those measured using a microscope objective. The data clearly demon-strate that microlens excitation consistently decreases the energy needed to reach the condensation thresholds.

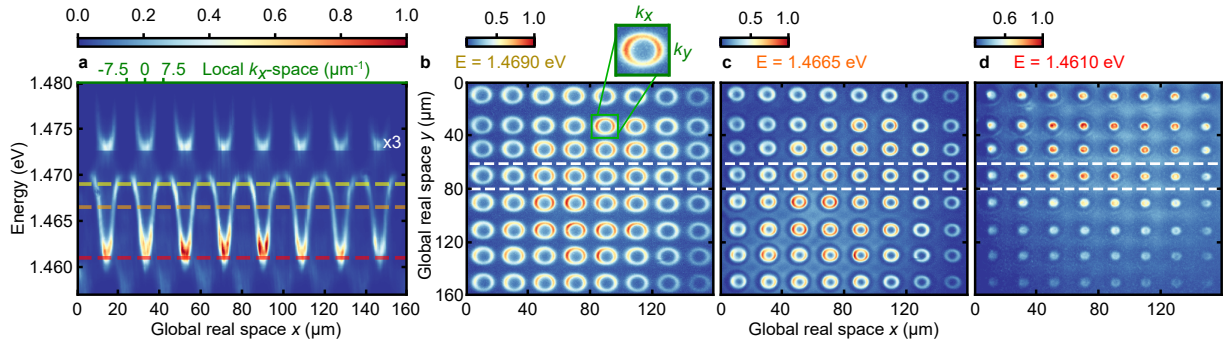

**Fig. 3 Simultaneous imaging of multiple exciton-polariton dispersion relations in a GaAs-based microcavity using an array of microlenses. a,** Momentum-resolved photoluminescence spectra of eight exciton-polariton dispersion branches collected from eight different positions on the sample using a configuration with eight vertically aligned ellipsoidal microlenses positioned along the spectrometer slit. The total signal was magnified and directed into the spectrometer using a microscope objective ($10\times$ magnification, NA = 0.30). The dashed lines on the map indicate the energies at which cross-sections of the dispersions were taken. **b–d,** Cross-sections at selected energies of the photoluminescence signal collected from an $8\times8$ microlens array, corresponding to the case shown in panel **(a)**. Each "circle" represents a separate momentum-space distribution from a distinct spatial position on the sample. The background of each map corresponds to the real-space emission from the sample surface. All measurements were performed at 4.5 K.

We attribute the reduced condensation threshold to the more efficient focusing provided by the microlens. The microlens exhibits a high numerical aperture (NA ≈ 1), offering a significantly smaller excitation spot compared to that produced by the microscope objective used in this measurements (NA = 0.55). Under diffraction-limited conditions, the microlens yields a spot area approximately three times smaller (2.1 $\mu m^2$ versus 0.6 $\mu m^2$). These results highlight that microlenses can serve not only as light-collection optics but also as efficient excitation elements, enhancing the local power density.

## Conclusions and future outlook

We replaced the conventional microscope objective in the detection path with printed microlenses, which serve as compact and highly efficient light-collection optical elements with numerical apertures reaching up to 0.95. This configuration enables wide-angle momentum-space imaging, even when used together with low-NA Bertrand optics. The microlens array allows for simultaneous momentum-space imaging across 64 independent spatial positions. Furthermore, we demonstrated that microlenses can also act as efficient excitation elements, as evidenced by the significantly reduced Bose-Einstein condensation thresholds in semiconductor optical microcavity, compared to those obtained using conventional cryogenic-compatible microscope objectives.

The ability to perform simultaneous momentum-space imaging of multiple spatial sites relying on the microlens array opens up, to the best of our knowledge, previously inaccessible experimental possibilities. Local and parallel access to multiple dispersion relations enables the study of spatially extended phenomena that change the distribution of interaction angles. These include, for example, parallel study of particle flows, relaxation dynamics, interactions across multiple sites, coupled oscillator phenomena, and non-Hermitian physics in momentum-space. Such experiments can be realized using the configuration demonstrated here and may be further extended by integrating streak cameras, where the projected signal forms a time-resolved, spatially multiplexed momentum-space image.

The samples used in this study are transmissive, which facilitates a straightforward separation of the excitation and detection paths. This geometry enhances experimental flexibility, enabling precise control over the spatial distribution of interacting polaritons and excitonic reservoir nodes that can be simultaneously probed in momentum-space. Both this approach and the use of microlenses for angle-resolved excitation are promising directions that we leave open for future studies.

Although this work focuses on exciton-polaritons in semiconductor microcavities, the advantages of printed microlens arrays extend far beyond this platform. Other photonic systems, such as waveguides[48,49], or monolayers of transition metal dichalcogenides[43], are compatible with direct-printing technologies and can fully exploit the benefits presented here for momentum-space imaging and beyond.

## Methods

### Two-photon lithography

Microlenses were fabricated using a *Photonic Professional* (Nanoscribe, GmbH) system based on the two-photon absorption process. This technique involves using a photosensitive resin that polymerizes only at the focal point of a tightly focused laser beam. The resin is applied on the sample surface, and then an immersion microscope objective with a numerical aperture (NA) greater than 1 is immersed directly into the resin. In our case, the refractive index of the IP-Dip resin was $n = 1.52$, and the immersion objective used had a numerical aperture of NA = 1.3.

The fabrication process utilized a Calmar Laser Mendocino fiber-coupled femtosecond laser with a wavelength of $\lambda = 785$ nm, a repetition rate of 50 MHz, and a pulse duration of approximately 100 fs. After selectively polymerizing the desired structures with the laser, the unpolymerized resin was removed by washing with solvents: di(propylene glycol) methyl ether and isopropanol. The structures were first rinsed in di(propylene glycol) methyl ether for 20 minutes and then in isopropanol for another 20 minutes, both times under magnetic stirring. Following the washing process, the structures were carefully dried.

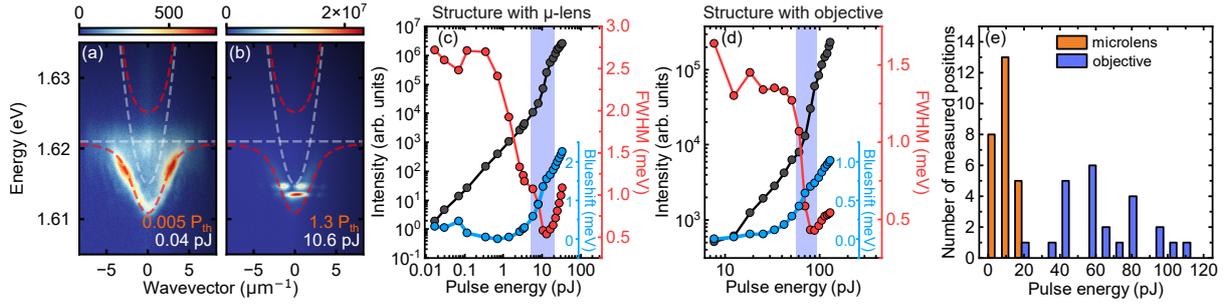

**Fig. 4 Bose-Einstein condensation of exciton-polaritons with a lowered threshold under excitation through an ellipsoidal microlens.** Momentum-resolved photoluminescence spectra from a CdTe-based microcavity with an ellipsoidal microlens printed on the cavity surface: **a**, below and **b**, above the condensation threshold. A $50\times$ microscope objective was used as the Bertrand lens. The excitation pulse energy and relative pulse energy are indicated in the bottom right corners. **c**, Analysis of the emission from the microcavity under microlens excitation with increasing laser pulse energy. The emission intensity (black), energy (blue), and full width at half maximum (red) were extracted from Lorentzian fits to cross-sections of the signal at $k = 0$. **d**, Analogous analysis performed for a condensate in the planar microcavity. **e**, Comparison of condensation thresholds for excitation through the microlenses (orange) and standard planar excitation using a microscope objective (royal blue) across 26 different positions. For two unpatterned positions, condensation was not achieved.

In Supplementary Information, section Images of the Printed Structures, SEM images of representative microlenses with different sizes fabricated using the two-photon polymerization process are shown in Figs. S3**a**–**d**, and an optical microscope image of white light reflection from an $8\times8$ microlens array is presented in Fig. S3**e**.

**Samples**

Sample A consisted of two Bragg reflectors (DBRs) was composed of five pairs of $SiO_2/TiO_2$ layers and a central cavity layer made of $SiO_2$. The structure was designed for a central wavelength of $\lambda_0 = 732.77$ nm, operating at room temperature without the need for placing in a cryostat.

Sample B is a microcavity made of two Bragg mirrors made of GaAs/AlAs. A lower mirror was composed of 15.5 pairs of GaAs/AlAs layers and an upper one composed of 12 pairs. Between DBRs, inside the cavity four stacks of $In_{0.05}Ga_{0.95}As$ quantum wells were placed. Each stack contained three 8 nm thick wells separated by 12 nm thick GaAs barriers. The structure was designed for a wavelength of $\lambda_0 = 860$ nm.

Sample C consisted of a CdTe-based microcavity containing six CdTe quantum wells, doped with manganese (each 20 nm thick) arranged in three stacks of two quantum wells, separated by 5 nm barriers. The Bragg mirrors were made of alternating layers of $Cd_{0.88}Zn_{0.04}Mg_{0.08}Te$ and $Cd_{0.40}Mg_{0.60}Te$, with the upper mirror containing 16 pairs and the lower mirror containing 20 pairs, and the designed wavelength was $\lambda_0 = 750$ nm.

For more details on all three samples and a comparison between experiment and simulation, see the SI, section Samples and Berreman Simulations.

**Experimental setups**

In the experiment, reciprocal space reflectance maps acquired for Sample A and Sample B were carried out using a halogen lamp. Photoluminescence measured for Samples B and C were performed using a Ti:sapphire Mira-900 laser operating in picosecond mode, with a pulse duration of approx. 4 ps and a repetition rate of 76 MHz. The excitation wavelength was adjusted depending on the sample being measured (for Sample B, it was 785 nm, corresponding to an energy of 1.58 eV, and for Sample C, 730 nm, corresponding to 1.70 eV), but excitation always occurred non-resonantly, at an energy corresponding to the first Bragg minima. Sample A was measured at room temperature, while the Samples B and C were placed in a cryostat. Unlike a standard reciprocal space imaging setup, with the microscope objective located outside the cryostat, in our setup this function was performed by a microlens printed directly on the sample surface, operating under cryogenic conditions. The variable element in our setup was a so-called Bertrand lens - the middle lens in the reciprocal space imaging system, which could be realized either with a conventional lens or a microscope objective. For more details, see the SI section Experimental Setups.

## Data availability

All data supporting the conclusions of this study are included in the article. The data presented in this study are available from the corresponding author upon reasonable request.

## Acknowledgments

We gratefully thank Zofia Dziekan and Mikołaj Rogóź for the assistance in operating the Nanoscribe system and the SEM microscope. We also thank Sławomir Ertman for his advice on the technological aspect of two-photon lithography. The authors also thank Benoit Deveaud for providing the GaAs optical microcavity sample. Finally, we acknowledge students Zuzanna Werner, Maria Popławska, and Jakub Lewandowski for their contributions to the student project, and Nina Mitroczuk for her support during optical measurements.

M.F. acknowledges the project No. 2023/49/N/ST3/03595, B.P. acknowledges the project No. 2020/37/B/ST3/01657, funded by the National Science Center, Poland. This work was supported by Quantum Optical Networks based on Exciton-polaritons - (Q-ONE) funding from the HORIZON-EIC-2022-PATHFINDER CHALLENGES EU programme under grant agreement No. 101115575.

## Author contributions

M.F., M.M., M.K. and B.P. conceived the idea; Ł.Z., A.B. made introduction to 3D microprinting and first test prints; M.F., M.M. made final microlens printing; M.F. and M.M. performed the optical experiments; P.O. developed the ray-tracing program with support of M.F.; P.O. performed Berreman simulation; W.P. grown a CdTe-based microcavity sample; M.F. analyzed and visualized experimental data; M.M. made visualization using Blender; M.F. made SEM images; M.F. and M.M. wrote the manuscript with input from all authors; B.P., P.W. and J.Sz. provided funding; B.P. and M.K. supervised the project.

## Competing interests

The authors declare no competing interests.

## Additional information

**Correspondence and requests for materials** should be addressed to B.P.

# Supplementary information: Multiplexed back focal plane imaging with on-chip integrated microlens array

M. Furman,[1] M. Muszyński,[1] P. Oliwa,[1] Ł. Zinkiewicz,[1] A. Bogucki,[1, 2]

J. Szczytko,[1] P. Wasylczyk,[1] W. Pacuski,[1] M. Król,[1] and B. Piętka[1, *]

[1]*Institute of Experimental Physics, Faculty of Physics,*

*University of Warsaw, ul. Pasteura 5, PL-02-093 Warsaw, Poland*

[2]*Center for Quantum Nanoscience, Institute for Basic Science,*

*Seoul 03760, Republic of Korea*

## RAY TRACING

### General information

In our approach, we use a custom algorithm written in Python. Each beam we consider is modeled as having zero beam area and zero divergence, meaning we neglect all effects related to, for example, the curvature of a lens across the finite size of the beam spot. Due to the $C_{\infty v}$ symmetry along the $z$-axis (the propagation axis), we consider propagation only within a specific plane, such as the $xz$-plane. However, the full 3D propagation can be easily reconstructed by rotating the data obtained for the $xz$-plane around the $z$-axis by an arbitrary angle $\varphi$.

In our analysis, each microlens is modeled as a section of an ellipse cut along an axis perpendicular to its major axis (parallel to its minor axis). The ellipse has two foci located along the $z$-axis at positions $(0, -c)$ and $(0, c)$, where $c = \sqrt{a^2 - b^2}$. The major axis of the ellipse, of length $2a$, lies along the $z$-axis, while the minor axis, of length $2b$, lies along the $x$-axis.

In our simulations, we consider ellipses approximately $\sim 10\,\mu$m in size, made of an organic polymer with a refractive index of $n = 1.52$. The ellipses have similar value of $a$ and $b$, but we vary the cutting position $d$ along the $z$-axis. The minimum value, $d = -a$, corresponds to the full ellipse, while $d = a$ corresponds to the complete removal of the lens (i.e., no ellipse is present). Therefore, $d \in [-a, a]$.

The microlens is placed on top of the DBR sample, so in all simulations we account for the effects of light propagation through the DBR layers. In general, for all types of simulations, we assume that the light source is located just below the inner layer of the top DBR.

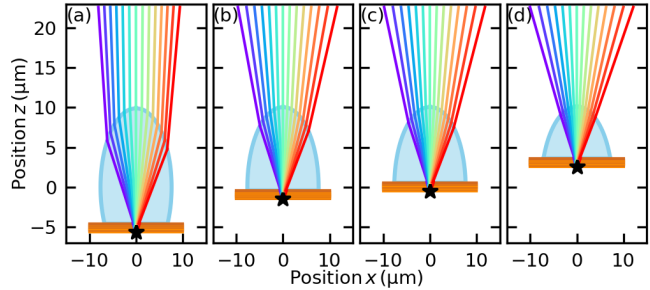

Figure S1. Beam propagation patterns for different emission angles at various values of $d$: (a) $d = -4.58\,\mu\mathrm{m}$, (b) $d = -0.44\,\mu\mathrm{m}$, (c) $d = 0.58\,\mu\mathrm{m}$, and (d) $d = 3.61\,\mu\mathrm{m}$. The black star marks the beam origin. The orange and brown rectangles represent the DBR layers, while the blue segment of the ellipse represents the microlens. The different colors of the propagating beams correspond to different emission angles.

In the first step, we use Snell's law to determine the beam's position after it passes through all the layers of the DBR – that is, at the top surface of the cavity. In the next step, we again apply Snell's law to calculate the propagation angle of the beam after it passes through the microlens. In the final step, depending on the specific case considered, we perform the final calculation of the beam's propagation in air.

A typical distribution of beam propagation for different angles and values of the parameter $d$ for SiO$_2$/TiO$_2$ cavity is shown in Fig. S1.

### Estimation of the Fourier plane shape

In the main manuscript, we present many measurements in reciprocal (Fourier) space. Therefore, the main goal of this simulation is to investigate the Fourier plane generated by the considered microlenses. In our approach, we estimate the Fourier plane using its standard geometric interpretation. Specifically, we consider several beams emitted at the same angle but from slightly different positions along the $x$-axis. Using the beam propagation method described in the previous section, we determine the points in space where these beams, emitted at the same angle but from different positions, intersect. These intersection points define locations in space where beams propagating at the same angle converge, according to standard geometric optics, these points belong to the Fourier plane. The full Fourier plane is reconstructed by considering a range of emission angles. In the next step, we performed such simulations for different values of the $d$ parameter. A typical distribution of the Fourier plane obtained for various microlenses is shown in Fig. S2. Figures S2(a) – S2(d) show the experimentally measured dispersion relations for microlenses with different values of $d$, while Figs. S2(e) – S2(h) present the corresponding simulated Fourier planes (teal curves).

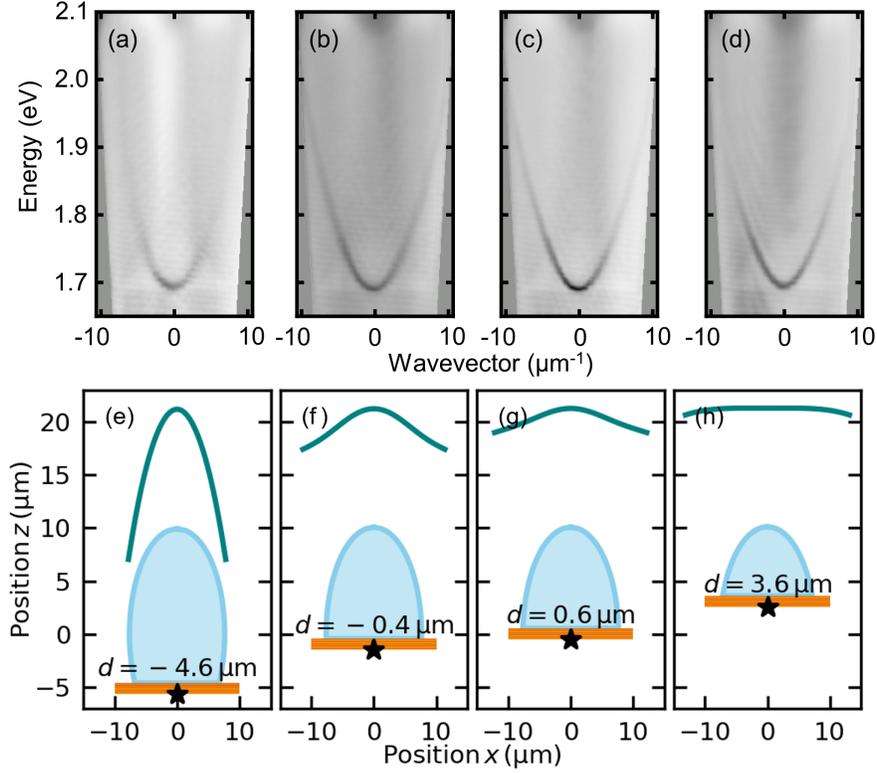

Figure S2. (a)–(d) Dispersion relations measured for the SiO$_2$/TiO$_2$ cavity with various printed microlenses. (e)–(h) Fourier planes (teal curves) obtained for different values of $d$ for microlenses printed on the SiO$_2$/TiO$_2$ cavity.

For small $d$ values ($d = -4.6\,\mu m$), only a limited angular range is observed. This effect results from the parabolic shape of the Fourier plane combined with the finite depth of field of the objective, which restricts the observable angular range. Naturally, adjusting the distance between the sample and the objective allows different parts of the dispersion relation to be observed.

Figures S2(b) and S2(c) show the dispersion relations for microlenses with optimized shapes. The corresponding simulated Fourier planes are shown in Figs. S2(f) and S2(g). In these cases, the Fourier plane can be well approximated by a Gaussian distribution, and due to the reduced variation along the $z$-axis, the entire plane is within the depth of field of the objective.

Table I. Parameters of microlenses printed on different samples.

| No. | Sample | $a$ [μm] | $b$ [μm] | $d$ [μm] |
|-----|--------|----------|----------|----------|
| 1 | SiO$_2$/TiO$_2$ | 9.92 | 7.62 | −4.58 |
| 2 | SiO$_2$/TiO$_2$ | 10.06 | 7.64 | −0.44 |
| 3 | SiO$_2$/TiO$_2$ | 10.07 | 7.65 | 0.58 |
| 4 | SiO$_2$/TiO$_2$ | 10.11 | 7.66 | 3.61 |
| 5 | CdTe | 10.28 | 7.95 | −0.21 |
| 6 | GaAs | 10.91 | 8.52 | 0.41 |

Finally, Figs. S2(d) and S2(h) present the case of a short microlens and its corresponding Fourier plane. Here, the Fourier plane is almost flat, but the observable angular range is significantly reduced compared to the optimized cases.

Based on these results, we selected the $d$ values to ensure that the Fourier plane resembles a Gaussian profile and that the microlens enables the observation of beams with the highest in-plane wavevectors. The microlens chosen for the measurements, along with the results obtained using it, is presented in the main manuscript and shown in Fig. S2(b) and (f). The parameters of this microlens correspond to those listed in row 2. of Table I. The parameters of this microlens, as well as those of the other microlenses, are summarized in Table I.

## IMAGES OF THE PRINTED STRUCTURES

The printed microlenses were subsequently imaged using a scanning electron microscope after depositing a thin gold layer on their surface [Fig. S3(a)–(d)]. Additionally, a photograph of the microlens array was taken within the experimental setup used later for optical measurements on the printed samples [Fig. S3(e)].

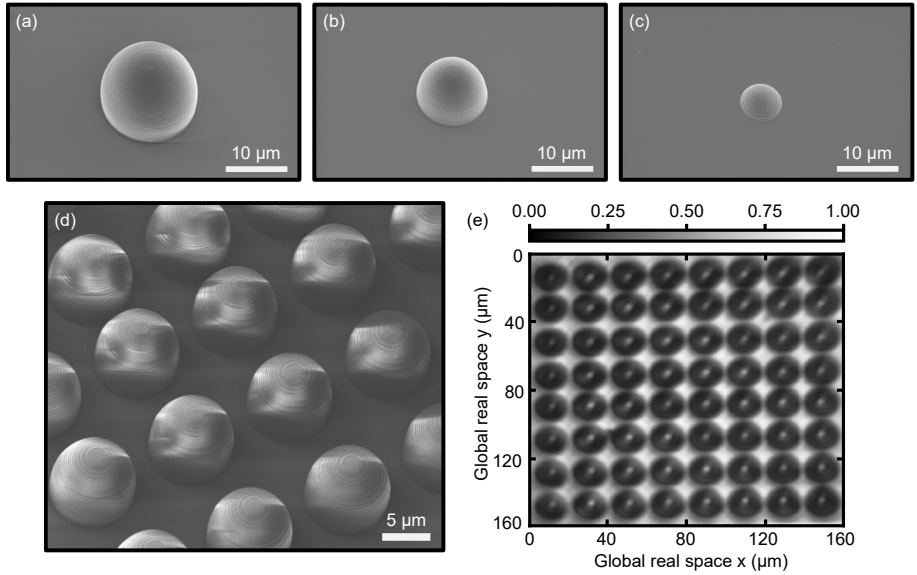

Figure S3. **SEM and optical images of microlenses printed using the two-photon polymerization process.** (a)–(c) SEM images of representative microlenses with varying sizes, (d) example SEM image of a microlens array taken at a 30-degree angle, (e) optical microscope image showing white light reflection from an 8×8 microlens array.

## EXPERIMENTAL SETUPS

The standard experimental setup used for back focal plane (BFP) imaging of semiconductor microcavity samples, shown in Fig. S4, was compatible with both room-temperature measurements and low-temperature experiments with the sample placed in a cryostat (as in the case of Sample B and Sample C).

The photoluminescence (PL) measurements of Sample B and Sample C were conducted in a reflective configuration. Excitation was provided by a Ti:Sapphire pulsed laser with a pulse duration of approx. 4 ps and 76 MHz repetition rate. Both Sample B and Sample C were excited non-resonantly, with the laser energy tuned to the first Bragg minimum. Sample B was excited using a laser with a wavelength of $\lambda_1 = 785$ nm (1.5794 eV), while Sample C was pumped with a laser of wavelength $\lambda_2 = 730$ nm (1.6984 eV). The laser power was controlled using a tunable neutral density filter. Next, the laser beam was directed to the beam splitter, which sends 50% of the power to a power meter, while the remaining portion was focused on the sample. To access momentum-space, we used either a conventional high NA, short working distance microscope objective (standard approach) or a printed microlens (proposed approach, see Fig. S5 for comparison) to focus the excitation beam on the sample and to collect the photoluminescence. The collected emission was then relayed to the detection system using a Bertrand lens and an imaging lens. The laser signal was filtered out using a spectral filter, and the emission from the sample was directed either to a CCD camera or to the entrance slit of a spectrometer. Additionally, the system included a halogen lamp that allowed for white-light reflection measurements from the microcavity and was used in reflectivity measurements of Sample A and Sample B.

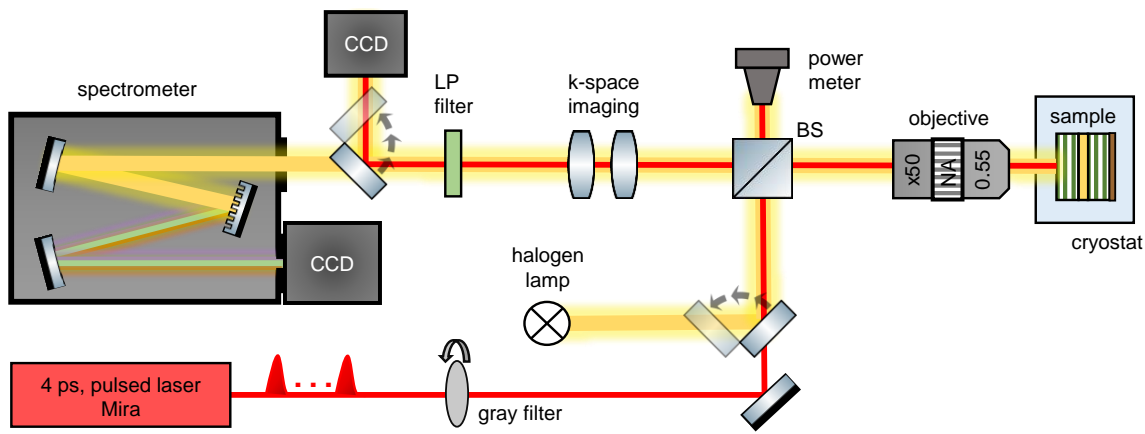

Figure S4. **Standard experimental setup for reciprocal space measurements of semiconductor optical microcavities, compatible with both room and cryogenic temperatures.**

Fig. S5 presents schematic diagrams of two experimental setups used for imaging the reciprocal space, illustrating the path of light rays as they pass through the components of each system. Fig. S5(a) shows a standard configuration employing a conventional microscope objective, which collects the emitted signal and forms the Fourier plane (FP) outside the cryostat. In contrast, Fig. S5(b) depicts a setup in which a microlens, placed directly on the sample surface, acting as the microscope objective. This configuration enables full collection of the emitted signal, with the Fourier plane forming approx. 10–15 μm above the microlens surface, still within the cryostat. In both setups, the remaining optical components, such as the Bertrand lens and the imaging lens relaying the signal to the spectrometer slit or CCD camera, are positioned outside the cryostat and can be freely adjusted as needed.

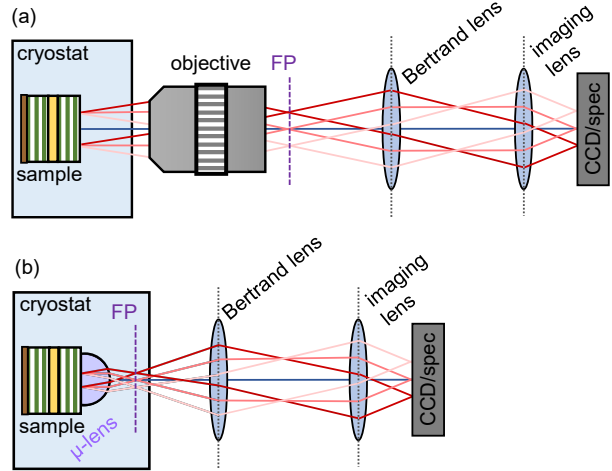

Figure S5. **Experimental setups for imaging reciprocal space, showing how light rays from the sample are directed through optical elements of the setup.** (a) Standard experimental setup with a microscope objective for collecting the signal from the sample, (b) experimental setup in which microlens play the role of the microscope objective.

## SAMPLES AND BERREMAN SIMULATIONS

### Sample A: SiO$_2$/TiO$_2$ cavity

This structure is a standard distributed Bragg reflector (DBR) cavity composed of two DBRs, each consisting of five pairs of SiO$_2$ and TiO$_2$ layers, resulting in a total of 20 layers. The inner cavity layer is made of SiO$_2$. The thicknesses of the TiO$_2$ and SiO$_2$ layers are given by $l_{TiO_2} = \lambda_0/4n_{TiO_2}(\lambda_0)$ and $l_{SiO_2} = \lambda_0/4n_{SiO_2}(\lambda_0)$, respectively, where $\lambda_0$ is the central wavelength of the cavity, $n_{TiO_2}(\lambda_0)$ is the refractive index of TiO$_2$ at $\lambda_0$, and $n_{SiO_2}(\lambda_0)$ is the refractive index of SiO$_2$ at the same wavelength. In the simulations, we assume $\lambda_0 = 732.77$ nm, and the dispersion relations for SiO$_2$ and TiO$_2$ are taken from Fig. S6.

### Sample B: GaAs microcavity

This structure is a DBR microcavity that contains quantum wells. The DBRs are made of GaAs and AlAs. The bottom DBR consists of 15.5 pairs of layers, while the top DBR consists of 12 pairs. The inner cavity layer is made of AlAs. The thickness of each layer is given by $l_{AlAs} = \lambda_0/4n_{AlAs}(\lambda_0)$ and $l_{GaAs} = \lambda_0/4n_{GaAs}(\lambda_0)$, where $\lambda_0 = 860$ nm. The refractive indices $n_{GaAs}(\lambda)$ and $n_{AlAs}(\lambda)$ as functions of wavelength are shown in Fig. S7.

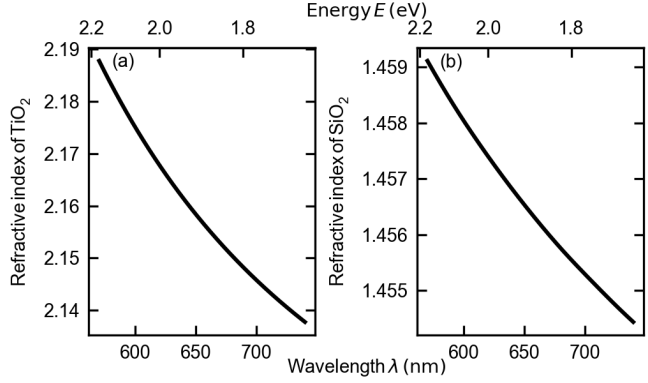

Figure S6. Refractive index as a function of wavelength for (a) TiO$_2$ and (b) SiO$_2$.

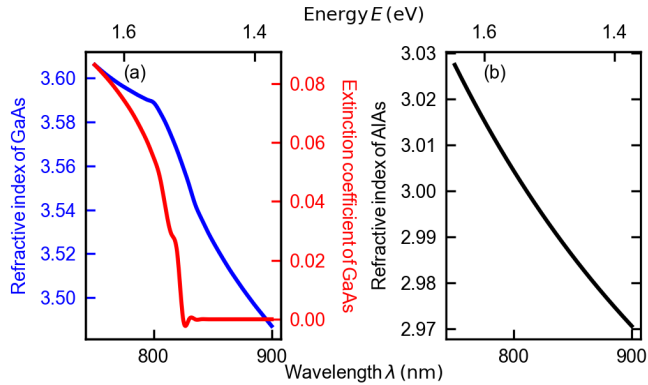

Figure S7. Refractive index as a function of wavelength for (a) GaAs and (b) AlAs. The blue and red curve in (a) panel denote the real and imaginary part of the complex refractive index, respectively.

In the considered wavelength range, GaAs exhibits strong absorption for wavelengths shorter than its bandgap wavelength ($\lambda_{GaAs} \approx 820$ nm). The refractive index data for GaAs are taken from [1], but we assume the extinction coefficient $\kappa$ is zero for $\lambda > 820$ nm, as no absorption is observed in our sample above this wavelength. Small variations in the value of $\kappa$ near $\lambda_{GaAs}$ arise from the interpolation algorithm.

The refractive index dependence for AlAs follows [2], since in the wavelength range of interest the refractive index of AlAs changes only slightly with temperature. Thus, assuming that the refractive index at helium temperature is very similar to that at room temperature

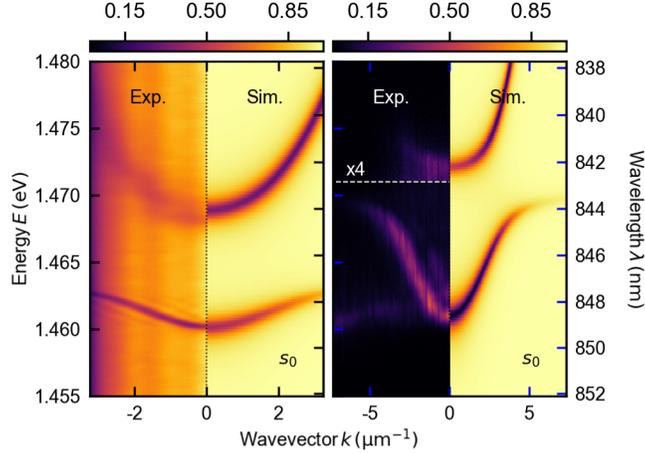

Figure S8. Measured reflectance and luminescence spectra compared with calculated reflectance spectra. Comparison for the GaAs cavity (a) near the microlens and (b) directly at the microlens. The luminescence signal for energies above $1.471\,\text{eV}$ was multiplied by a factor of 4 to better visualize the upper polariton branch.

is sufficient to accurately reproduce the experimental data.

This cavity contains four stacks of $\text{In}_{0.05}\text{Ga}_{0.95}\text{As}$ quantum wells. Each stack consists of three 8 nm quantum wells separated by 12 nm GaAs barriers. Each stack is separated from adjacent stacks or from the top/bottom DBRs by a GaAs layer approximately 80 nm thick.

The excitonic response in the $\text{In}_{0.05}\text{Ga}_{0.95}\text{As}$ quantum wells can be well described using the Lorentz model:

$$\varepsilon = \varepsilon_\infty + \frac{f}{E^2 - E_X^2 - i\gamma E} \tag{1}$$

where $f$ denotes the oscillator strength, $E_X$ is the exciton energy, $\gamma$ is the damping coefficient, and $\varepsilon_\infty$ is the dielectric constant at infinite energy. The complex refractive index in this case is given by $n = \sqrt{\varepsilon}$.

For this quantum well, we assume $\varepsilon_\infty = 3.5^2$, $f = 9 \times 10^{-3}\,(\text{eV})^2$, $\gamma = 8 \times 10^{-4}\,\text{eV}$, and exciton energies $E_{X,c} = 1.464\,\text{eV}$ for the planar cavity and $E_{X,\mu} = 1.470\,\text{eV}$ for the cavity with a microlens. In the GaAs cavity, a significant shift in the exciton energy is observed due to strain induced by the microlens.

We determined the parameters of the Lorentz model by comparing the reflectivity spectra measured near the microlens with the reflectivity spectra obtained from Berreman calculations. This comparison is shown in Fig. S8(a). Next, we compared the measured luminescence spectra collected at the microlens with the reflectivity spectra calculated using

Berreman method, but with the updated exciton energy $E_{X,\mu}$. Following the same procedure as previously, we establish the relation between the CCD camera pixels in the spectrometer and the in-plane wavevector. The comparison between the luminescence and reflectivity spectra is shown in Fig. S8(b). To better visualize the upper polariton branch, we multiplied the luminescence signal for energies greater than 1.471 eV by a factor of 4.

In this case, the estimated numerical aperture of investigated microlens is approximately 0.95.

## Sample C: CdTe cavity

This structure is a DBR microcavity that contains quantum wells. The DBRs are composed of $Cd_{0.88}Zn_{0.04}Mg_{0.08}Te$, referred to as the first material, with a refractive index of $n_1 = 2.74$, and $Cd_{0.40}Mg_{0.60}Te$, referred to as the second material, with a refractive index of $n_2 = 2.46$. The top mirror consists of 16 pairs of layers, while the bottom mirror contains 20 pairs. The thickness of each layer in the DBRs is $l_1 = \lambda_0/4n_1$ or $l_2 = \lambda_0/4n_2$, where $\lambda_0 = 750$ nm is the central wavelength. The inner cavity layer is made of the second material.

The cavity contains six CdTe quantum wells, each with a thickness of $l_q = 20$ nm, doped with manganese and arranged in three stacks. Each stack consists of two quantum wells separated by a $l_b = 5$ nm barrier made of the first material. The stacks themselves, as well as the area between the outer stacks and the top or bottom DBR, are separated by the first material with a thickness of approximately $l_s \approx 114$ nm.

Similarly to the GaAs quantum wells, the refractive index dependence of the CdTe quantum wells is described by the Lorentz model [see, Eq. (1)].

For this quantum well, we assume the following parameters: $\varepsilon_\infty = 2.7^2$, $f = 1.5 \times 10^{-2}$ (eV)$^2$, $\gamma = 5 \times 10^{-3}$ eV, and the exciton energy $E_{X,c} = 1.620$ eV for the planar cavity, and $E_{X,\mu} = 1.621$ eV in the region with the microlens. This small shift in exciton energy is attributed to strain induced by the microlens.

The parameters in the Lorentz model were chosen by comparing the reflectivity spectra calculated using the Berreman method with the experimental luminescence spectra collected near the investigated microlens. The comparison between these two data sets is shown in Fig. S9(a).

In the next step, we used Berreman method to calculate the reflectance spectra for a similar cavity, but with the exciton energy shifted to $E_{X,\mu}$. By comparing the luminescence spectra collected for the microlens with the reflectance spectra obtained from the Berreman calculations, we determined the conversion factor between the pixel position on the CCD camera in the spectrometer and the in-plane wavevector. The comparison between luminescence spectra and the results of the Berreman calculation is shown in Fig. S9(b).

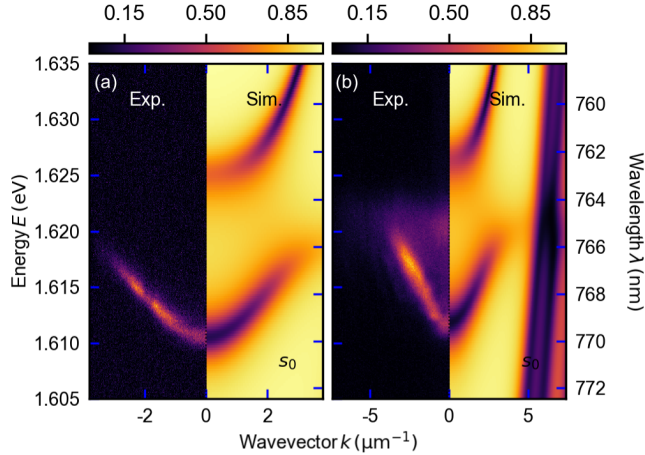

Figure S9. Measured luminescence spectra (Exp.) and simulated reflectance spectra obtained from Berreman calculations (Sim.). Comparison for the CdTe cavity (a) near the microlens and (b) directly at the microlens.

In this case, the estimated numerical aperture of the investigated microlens is approximately 1.0.

## ADDITIONAL EXPERIMENTAL RESULTS

### The coupled oscillator model

In order to further analyze the polariton dispersion, we successfully fitted a coupled oscillator model to the experimental data points obtained from Lorentzian fits of the dispersion maps. This procedure was performed for both detection configurations, using the microscope objective and the microlens, and for both investigated samples: Sample B and Sample C. The coupled oscillator model provides a good agreement with the extracted energy maxima, confirming the strong coupling regime between excitons and photons in Sample B and Sample C microcavity structures.

The Hamiltonian describing the strongly coupled polariton system is given by:

$$H = \begin{pmatrix} E_{\text{ph}}(k) & \frac{\hbar\Omega}{2} \\ \frac{\hbar\Omega}{2} & E_{\text{ex}}(k) \end{pmatrix} \qquad (2)$$

where $E_{\text{ph}}$ and $E_{\text{ex}}$ represent the bare photon and exciton energies, $\Omega$ is the Rabi frequency, respectively, and $\hbar$ is the reduced Planck constant. The eigenvalues of this Hamiltonian, corresponding to the energies of the lower and upper polariton branches, are given by:

$$E_{\text{LP,UP}} = \frac{E_{\text{ph}} + E_{\text{ex}}}{2} \mp \frac{1}{2}\sqrt{(E_{\text{ph}} - E_{\text{ex}})^2 + \hbar^2\Omega^2} \qquad (3)$$

These expressions were used to fit the experimental data, allowing extraction of key parameters such as the Rabi splitting $\hbar\Omega$ and the bare mode energies.

**Comparison of the angular range for dispersion with a microscope objective with NA = 0.55 and with a microlens with NA= 0.95.**

To demonstrate that using a microlens with a numerical aperture NA = 0.95, which collects more light rays and collimates them into a beam, enables access to a wider angular range, and consequently a broader range of the in-plane wavevector, we compared the dispersion obtained from a planar microcavity structure collected using a microscope objective with NA = 0.55 [Fig. S10(a)], to that obtained using the microlens, as shown in Fig. S10(c).

For each dispersion map, cross-sections were taken at selected wavevector values, and Lorentzian fits were fitted to the resonant energies of resulting spectra. From these fits, the energy peak positions were extracted and plotted as a function of the in-plane wavevector in Fig. S10(b) for the microcavity with the microscope objective, and in Fig. S10(d) for the microlens case. The accessible NA ranges are indicated on Fig. S10(b) and (d) using dashed lines, orange for NA = 0.55 and blue for NA = 0.95.

To minimize inconsistencies in polariton detuning caused by printing-induced strain, the y-axis in both spectra is presented as 'Relative energy,' with the zero-energy reference set to the minimum of the lower polariton branch in each case.

The fitting results clearly confirm that the use of a microlens extends the accessible in-plane wavevector range of the detected signal and allows full observation of the polariton dispersion relation across the entire wavevector range corresponding to the microlens

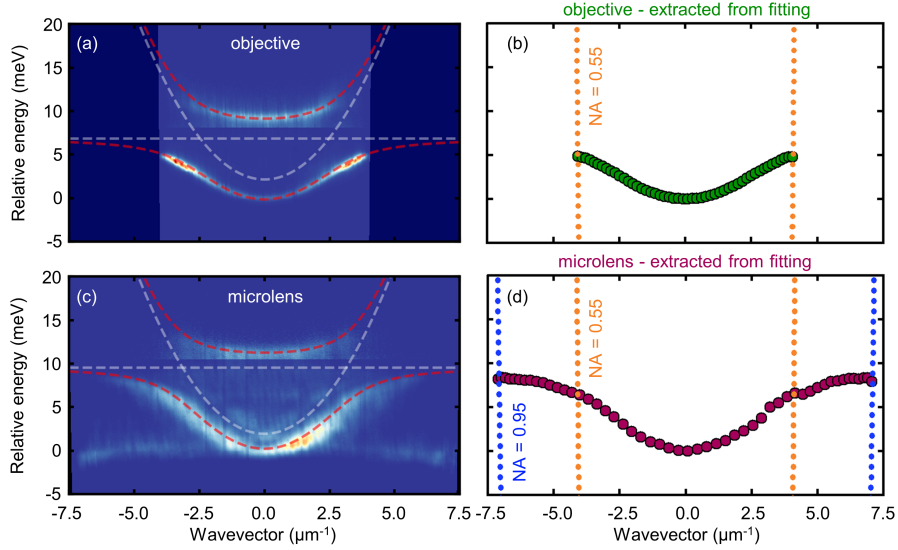

Figure S10. **Comparison of dispersion for Sample B from the planar microcavity and the microcavity structure with an imprinted microlens.** (a) Photoluminescence map in reciprocal space obtained from the planar structure using an objective with NA = 0.55, accompanied by a fitted coupled oscillator model. (b) Emission maxima extracted via a Lorentzian fit to the lower polariton branch in panel (a). The orange dashed lines indicate the collection range defined by the numerical aperture of the objective. (c) Photoluminescence map of the structure incorporating the imprinted microlens, along with the corresponding fitted model. (d) Lorentzian fit of the emission maxima for the data presented in panel (c). Orange dashed lines indicate the NA = 0.55 range for the microscope objective and blue dashed lines of the NA = 0.95 range of the microlens, shown for comparison.

numerical aperture.

## Simultaneous momentum-space imaging with lens

Fig. S11 presents analogues results to those shown in Fig. 3 from the main manuscript, but obtained using a different optical configuration. In this setup, the role of the second lens in the experimental system (commonly referred to as the Bertrand lens), which is typically placed after the microscope objective, was fulfilled by a single lens with a short focal length ($f = 25.4$ mm), instead of a microscope objective as in Fig. 3. These measurements confirm that reciprocal space imaging is possible in the microlens-based configuration using only low-complexity optics, without the need for a conventional microscope objective.

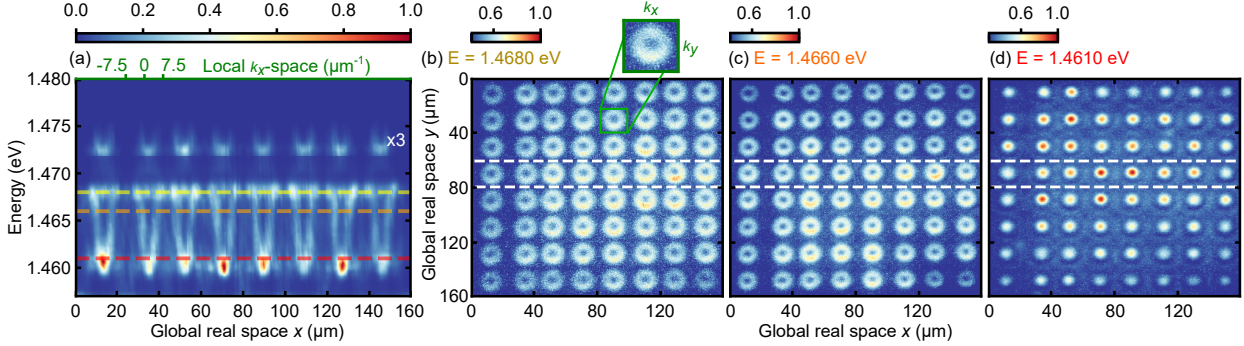

Figure S11. **Simultaneous imaging of multiple exciton-polariton dispersion relations in a GaAs-based microcavity using microlenses where a short-focus ($f = 25.4$ mm) lens serves as the Bertrand lens.** Panel (a) shows momentum-resolved photoluminescence spectra from eight exciton-polariton dispersion branches collected at eight different positions on the sample using a configuration with eight vertically aligned ellipsoidal microlenses positioned along the spectrometer slit. The signal was magnified using a convex lens with a focal length of $f = 25.4$ mm, acting as a Bertrand lens. Dashed lines on both maps indicate the energies at which the dispersion cross-sections were taken. (b)–(d) Cross-sections at selected energies of the photoluminescence signal collected from an 8×8 microlens array, corresponding to the case of the map in panel (a). Each visible "circle" represents a separate momentum-space distribution from a separate position on the sample. The background of each map contains real space emission from the sample surface. All measurements were made at 4.5 K.

The resolution achieved in this setup is lower than that obtained with a microscope objective, due to the lens limited correction of optical aberrations such as field curvature and spherical aberration, but the reciprocal space cross-sections in the $k_x$–$k_y$ plane at selected energies can still be clearly visualized.

### Condensation for representative dispersions at second detuning

In the main text of the manuscript, Fig. 4 (e) presents the condensation threshold statistics for a CdTe-based microcavity. The threshold values were extracted from two different regions of the sample, each characterized by a different detuning between the photonic and excitonic modes. The dispersion maps and signal characteristics from both the microlens and the planar structure corresponding to the first detuning are shown in Fig. 4 (a,b). The analogous data for the second detuning, the dispersion maps and emission characteristics are presented in Fig. S12.

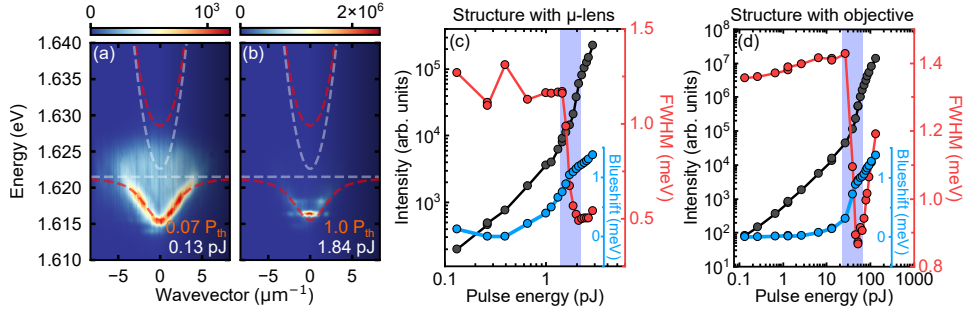

Figure S12. **Bose–Einstein condensation of exciton-polaritons with a lowered threshold under excitation through an ellipsoidal microlens for the second detuning from the Fig. 4(e), measured in the CdTe-based microcavity.** (a) Momentum-resolved PL map below the condensation threshold under microlens excitation, and (b) at the threshold under microlens excitation, (c) signal intensity at $k = 0$ collected through the microlens and (d) signal intensity at $k = 0$ from the planar region of the structure.

---

\* barbara.pietka@fuw.edu.pl